\documentclass[prd,twocolumn,a4paper,superscriptaddress,floatfix]{revtex4}
\usepackage{graphicx}
\usepackage{bbm}
\usepackage{amssymb}
\usepackage{calligra}
\usepackage{lipsum}
\usepackage[T1]{fontenc}
\DeclareFontShape{T1}{calligra}{m}{n}{<->s*[2.5]callig15}{}
\DeclareMathAlphabet{\mathcalligra}{T1}{calligra}{m}{n}
\DeclareMathAlphabet{\mathpzc}{OT1}{pzc}{m}{it}

\begin{document}
%My commands
\newcommand{\be}{\begin{equation}}
\newcommand{\ee}{\end{equation}}
\newcommand{\bq}{\begin{eqnarray}}
\newcommand{\eq}{\end{eqnarray}}
\newcommand{\bsq}{\begin{subequations}}
\newcommand{\esq}{\end{subequations}}
\newcommand{\bc}{\begin{center}}
\newcommand{\ec}{\end{center}}
\newcommand\lapp{\mathrel{\rlap{\lower4pt\hbox{\hskip1pt$\sim$}} \raise1pt\hbox{$<$}}}
\newcommand\gapp{\mathrel{\rlap{\lower4pt\hbox{\hskip1pt$\sim$}} \raise1pt\hbox{$>$}}}
\newcommand{\dpar}[2]{\frac{\partial #1}{\partial #2}}
\newcommand{\sdp}[2]{\frac{\partial ^2 #1}{\partial #2 ^2}}
\newcommand{\dtot}[2]{\frac{d #1}{d #2}}
\newcommand{\sdt}[2]{\frac{d ^2 #1}{d #2 ^2}}
\newcommand{\vv}[0]{{\bar v}}
\newcommand{\ave}[1]{\left< #1 \right>}

\title{Stochastic Gravitational Wave Background generated by Cosmic String Networks: Velocity-Dependent One-Scale model versus Scale-Invariant Evolution}

\author{L. Sousa}
\email[Electronic address: ]{Lara.Sousa@astro.up.pt}
\affiliation{Centro de Astrof\'{\i}sica da Universidade do Porto, Rua das Estrelas, 4150-762 Porto, Portugal}
\affiliation{Departamento de F\'{\i}sica e Astronomia da Faculdade de Ci\^encias
da Universidade do Porto, Rua do Campo Alegre 687, 4169-007 Porto, Portugal}

\author{P.P. Avelino}
\email[Electronic address: ]{ppavelin@fc.up.pt}
\affiliation{Centro de Astrof\'{\i}sica da Universidade do Porto, Rua das Estrelas, 4150-762 Porto, Portugal}
\affiliation{Departamento de F\'{\i}sica e Astronomia da Faculdade de Ci\^encias
da Universidade do Porto, Rua do Campo Alegre 687, 4169-007 Porto, Portugal}

\begin{abstract}
We compute the power spectrum of the stochastic gravitational wave background generated by cosmic string networks described by the Velocity-Dependent One-Scale (VOS) model, for a wide range of macroscopic and microscopic parameters. The VOS model --- which has been shown to provide an accurate macroscopic description of the evolution of cosmic string networks --- is used to demonstrate that cosmic string networks are unable to rapidly attain scale-invariant evolution after the transition between the radiation and matter eras. However, in computations of the stochastic gravitational wave background, it is often assumed that the networks experience scale-invariant evolution throughout cosmological history. We demonstrate that this assumption leads to an underestimation of the amplitude and broadness of the peak of the spectrum, that may consequently lead to inaccurate observational constraints on the cosmic string tension.
\end{abstract} 
\maketitle

\section{Introduction}

The production of cosmic string networks as remnants of symmetry breaking phase transitions in the early universe is predicted in a wide variety of grand unified scenarios. Initial attempts to model cosmic string networks \cite{Kibble:1984hp,Bennett:1985qt,Bennett:1986zn} were based on the assumption that a single lengthscale is sufficient to describe their dynamics. These one-scale models were later ameliorated by treating the average root-mean-square velocity of the network as a dynamical variable, while maintaining the one-scale assumption. This Velocity dependent One-Scale (VOS) model \cite{Martins:1996jp,Martins:2000cs} provides a quantitative description of the string network throughout its evolution and it has been thoroughly tested and calibrated using numerical simulations of cosmic string network dynamics \cite{Bennett:1990uza,Allen:1990tv,Moore:2001px,Martins:2003vd}. For a generalization of the VOS model for networks of domain walls or $p$-brane of arbitrary dimensionality, see \cite{Avelino:2005kn,Sousa:2010zza,Avelino:2010qf,Avelino:2011ev,Sousa:2011ew,Sousa:2011iu}.

Cosmic string interactions are crucial to the dynamics of cosmic string networks. These interactions result in the formation of closed cosmic string loops that detach from the long string network. Cosmic string loops oscillate under the effect of their tension and decay by emitting gravitational waves. %In this process, larger loops may also break into smaller loops which have a faster decay into gravitational radiation than their progenitors.
This emission occurs throughout the cosmological history and gives rise to a Stochastic Gravitational wave background (SGWB) with a broad frequency range \cite{Vilenkin:1981bx,Hogan:1984is,Brandenberger:1986xn,Accetta:1988bg}. Different regions of the stochastic gravitational wave background power spectrum may be probed using current and upcoming astrophysical experiments: direct gravitational wave detectors --- either ground-based (Advanced LIGO \cite{Sigg:2008zz}, Advanced Virgo \cite{Accadia:2011zzc} and KAGRA \cite{Kuroda:2010zzb}) or space-borne (evolved LISA/NGO \cite{AmaroSeoane:2012km} and DECIGO \cite{Kawamura:2011zz}); pulsar timing experiments (Parkes \cite{Manchester:2007mx} and European \cite{Ferdman:2010xq} pulsar timing arrays and NANOGRAV \cite{Demorest:2012bv}); small-scale fluctuations and B-mode polarization of Cosmic Microwave Background \cite{Smith:2006nka} (Planck \cite{Planck:2006aa} and CMBpol \cite{Baumann:2008aq}); and big bang nucleosynthesis \cite{Cyburt:2004yc}. There is thus the prospect either for the detection of the SGWB generated by cosmic string networks or for the tightening of current constraints on the cosmic string tension.

The shape and amplitude of the stochastic gravitational wave background power spectrum is highly dependent on both the large-scale properties of the network and the size and emission spectrum of the cosmic string loops. Previous estimations and computations of the SGWB spectrum \cite{Caldwell:1991jj,Damour:2001bk,Damour:2004kw,Siemens:2006vk,Hogan:2006we,DePies:2007bm,Olmez:2010bi,Binetruy:2012ze,Kuroyanagi:2012wm,Sanidas:2012ee} generated by cosmic string networks were performed using the one-scale model, thus implicitly assuming that the networks experience scale invariant evolution throughout cosmological history, and that the scaling parameters suffer a sudden change at the radiation-matter transition. This, however, does not give an accurate picture of network evolution: during the transition between the radiation- and matter-dominated eras, the network adapts slowly to the changes in the underlying background dynamics. In this paper, we study the SGWB generated by cosmic string networks described by the VOS model. We will find that the assumption of scale-invariant evolution during the radiation- and matter-dominated eras has indeed a significant impact on the shape and amplitude of the spectra and may, thus, result in inaccurate constraints on the cosmic string tension.

This paper is organized as follows. In Sec. \ref{string}, we review the VOS model for cosmic string network dynamics, and provide a brief description of network evolution throughout cosmological history. In Sec. \ref{SGWB}, we describe the emission of gravitational waves by cosmic string loops, and present a method for computing the SGWB power spectrum that results from the superimposition of the individual emissions by many cosmic string loops. In Sec. \ref{transition}, we compute the stochastic gravitational wave background generated by cosmic string networks for a wide variety of parameters and analyze the effect of considering a realistic radiation-matter transition on the shape and amplitude of the spectrum. We then conclude in Sec. \ref{con}.

\section{Microscopic Cosmic String Network evolution\label{string}}

The world-history of an infinitely thin and featureless cosmic string may be described by a two-dimensional worldsheet,
\be
x^\sigma=x^\sigma(u^a)\,,
\ee
obeying the Nambu-Goto action
\be
S=-\mu\int d^2 u\sqrt{\left|{\tilde g}\right|}\,,
\label{action}
\ee
where $a=0,1$, $u^0$ ($u^1$) are timelike (spacelike) coordinates that parameterize the cosmic string worldsheet, and $\mu$ is the cosmic string tension. Here, ${\tilde g}=\det ({\tilde g}_{\sigma\nu})$, ${\tilde g}_{\sigma\nu}=g_{\sigma\nu}x^\sigma_{,a}x^\nu_{,b}$ is the induced metric, and $g_{\sigma\nu}$ is the metric tensor of the background. In a flat $3+1$-dimensional homogeneous and isotropic Friedmann-Robertson-Walker (FRW) background, the line element is given by
\be
d^2 s=g_{\sigma\nu}dx^{\sigma}dx^{\nu}=a^2(\eta)\left(d^2\eta - d{\bf x}\cdot d{\bf x}\right)\,,
\ee
where $a$ represents the cosmological scale factor, $t$ and $\eta=\int dt/a$ are respectively the physical and conformal times, and ${\bf x}$ is a $3$-vector whose components are comoving cartesian coordinates.

By varying the action in Eq. (\ref{action}) with respect to $x^\sigma$, and imposing temporal-transverse gauge conditions,
\be
u^0=\eta\,, \qquad \dot{\bf x}\cdot {\bf x}'=0\,,
\label{gauge}
\ee
the string equations of motion can be written as \cite{Turok:1984db}
\bq
\ddot{\bf x}+2\mathcal{H}\left(1-\dot{\bf x}^2\right)\dot{\bf x} & = & \frac{1}{\epsilon}\left(\frac{{\bf x}'}{\epsilon}\right)'\label{ng1}\,,\\
\dot{\epsilon}+2\mathcal{H}\dot{\bf x}^2\epsilon & = & 0\label{ng2}\,,
\eq
where 
\be
\epsilon=\left(\frac{{\bf x}'^2}{1-\dot{\bf x}^2}\right)^{1/2}\,,
\ee
$\mathcal{H}=\dot{a}/a=Ha$, $H$ is the Hubble parameter, and dots and primes represent partial derivatives with respect to the conformal time and to the spacelike parameter $u\equiv u^1$, respectively. 

\subsection{Velocity-Dependent One-Scale Model\label{vos}}

The cosmic string equations of motion (Eqs. (\ref{ng1}) and (\ref{ng2})) can be averaged in order to obtain evolution equations which describe statistically the large-scale evolution of cosmic string networks. This might be done by assuming that the network is roughly homogeneous on sufficiently large scales --- and thus it may be described by an unique lengthscale --- and by treating the root-mean-square (RMS) velocity of the network as a dynamical variable. This model --- the Velocity-Dependent One-scale (VOS) Model \cite{Martins:1996jp,Martins:2000cs} --- provides a quantitative description of the large-scale evolution of cosmic string networks from early to late cosmological times (see Refs. \cite{Avgoustidis:2007aa,Sousa:2011ew,Sousa:2011iu} for a more general approach).

The total energy in cosmic strings, $E$, and the RMS velocity of the string network, $\vv$, are defined respectively as
\bq
E=\mu a(\eta)\int \epsilon du\label{energy}\,,\\
\vv^2\equiv\ave{\dot{\bf x}^2}=\frac{\int \dot{\bf x}^2 \epsilon du}{\int \epsilon du}\label{rms}\,.
\eq
The average string energy density, $\rho$, is proportional to $E a^{-3}$. By using Eqs. (\ref{energy}) and (\ref{ng1}-\ref{ng2}) and neglecting, for the moment, the energy loss caused by loop production, one obtains
\be
\frac{d \rho}{dt}+2H\rho(1+\vv^2)=0\,.
\label{rho-evo}
\ee

Cosmic string interactions are a key ingredient in the evolution of cosmic string networks. When two cosmic strings collide, they may exchange partners and intercommute. This process has important consequences for the dynamics of a cosmic string network, since it leads to the generation of cosmic string loops which detach from the long-string network. These loops have a finite lifespan, leading to an energy loss by the string network. A cosmic string network then has mainly two constituents: long strings that stretch over large scales (larger than the horizon), and smaller cosmic string loops. The large-scale properties of the long string network may be described by an unique scale --- the characteristic length $L$ --- defined as
\be
\rho \equiv \frac{\mu}{L^2}\,,
\label{L-def}
\ee
where $\rho$ is the long string energy density. The rate of energy loss caused by loop production may be written as \cite{Kibble:1984hp}
\be
\left.\frac{d\rho}{dt}\right|_{\rm loops}={\tilde c}{\bar v}\frac{\rho}{L}\,,
\label{loss}
\ee
where ${\tilde c}$ is a phenomenological parameter which characterizes the efficiency of this energy loss mechanism (this term needs to be added to Eq.  (\ref{rho-evo})). Numerical simulations indicate that, for ordinary cosmic strings, ${\tilde c}=0.23\pm 0.04$ is a good fit both in the matter and radiation eras \cite{Martins:2000cs}.

Using Eqs. (\ref{rho-evo})-(\ref{loss}), one may find the evolution equation for the characteristic length, $L$,
\be
2\frac{dL}{dt}=\left(2H+\frac{{\bar v}^2}{\ell_d}\right)L+{\tilde c}\vv\label{vos-L}\,.
\ee
Here, we have introduced the damping lengthscale,
\be
\ell_d^{-1}=2H+\ell_f^{-1}\,,
\ee
which accounts, not only for the damping caused by the background expansion, but also for the effect of frictional forces caused by the interaction of cosmic strings with other cosmological components (encoded in the friction lengthscale $\ell_f$). Throughout this paper, we shall assume that $\ell_f=\infty$, so that 
$\ell_d^{-1}=2H$.

The evolution equation for $\vv$ might be obtained by differentiating Eq.(\ref{rms}) and using Eqs. (\ref{ng1}-\ref{ng2})
\be
\frac{d\vv}{dt}=\left(1-\vv^2\right)\left[\frac{k}{L}-\frac{\vv}{\ell_d}\right]\label{vos-v}\,.
\ee
Here $k$ is a dimensionless curvature parameter that depends, in general, on $\vv$ (for a detailed definition of $k$ see \cite{Martins:1996jp,Sousa:2011ew}). In \cite{Martins:2000cs}, the following ansatz was suggested
\be
k(\vv)=\frac{2\sqrt{2}}{\pi}\left(1-\vv^2\right)\left(1+2\sqrt{2}\vv^3\right)\frac{1-8\vv^6}{1+8\vv^6}\,.
\ee
Note that the energy loss due to loop production is included implicitly in Eq. (\ref{vos-v}) through the dependency of the curvature term on $L$.

In order to also take into account the effect of the gravitational back-reaction on the dynamics of the long string network, one needs to add an extra term to the right-hand side of Eq. (\ref{vos-L}):
\be
\left(\frac{dL}{dt}\right)_{gr}=4{\tilde \Gamma} G\mu {\bar v}^6\,,
\ee
where ${\tilde \Gamma}\sim 65$ \cite{Vilenkin:1981bx,Quashnock:1990wv,Martins:2000cs} is a parameter which describes the efficiency of the emission of gravitational waves by long strings.

%%%

\subsection{The Scale-Invariant Regime\label{lin}}

It has been demonstrated numerically \cite{Bennett:1987vf,Albrecht:1989mk,Allen:1990tv} and using a variety of semi-analytical models \cite{Copeland:1991kz,Vincent:1996rb,Martins:1996jp,Martins:2000cs} that cosmic string networks may evolve towards a linear scaling regime during which the characteristic length grows proportionally to the particle horizon. In such conditions, the string energy density remains a constant fraction of the background energy density in both matter- and radiation-dominated epochs. Cosmic strings networks are thus not expected --- unlike domain walls and magnetic monopoles --- to dominate the energy density of the universe at late times.

A linear scaling regime of the form
\be
L=\xi t \qquad\mbox{and}\qquad\vv=\mbox{constant}\,,
\label{sca-def}
\ee
with
\be
\xi=\sqrt{\frac{k(k+{\tilde c})}{4\beta(1-\beta)}}\,,\quad\vv=\sqrt{\frac{k}{k+{\tilde c}}\frac{1-\beta}{\beta}}\,,
\label{sca-c}
\ee
is an attractor solution of the VOS equations (Eqs (\ref{vos-L}-\ref{vos-v})), in the case of decelerating power-law expansion of the universe --- with $a\propto t^\beta$ and $0<\beta<1$. (For a more detailed discussion of the scaling solutions of cosmic string networks, see \cite{Sousa:2011ew,Sousa:2011iu,Avelino:2012qy}.)

\begin{figure}
\begin{center}
\includegraphics[width=3.3in]{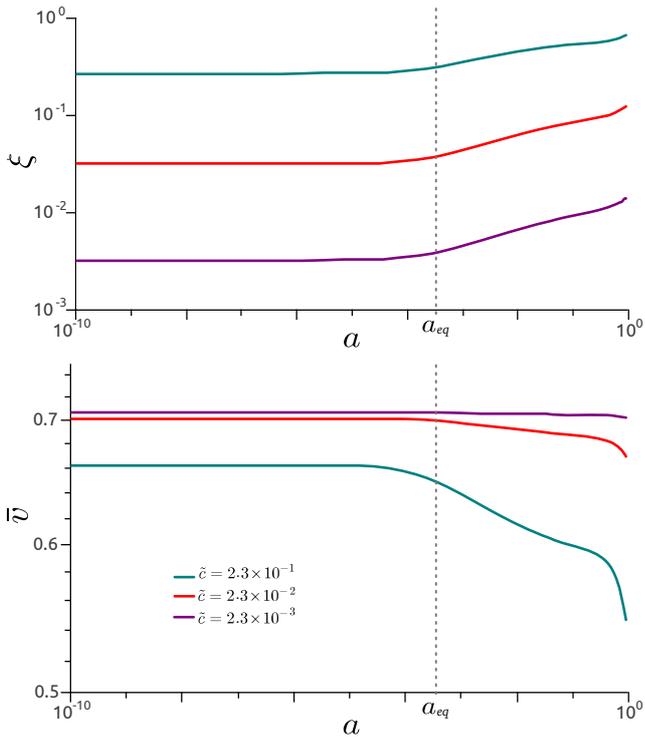}
\caption{Evolution of $\xi$ (top panel) and of the RMS velocity $\bar v$ (lower panel) of a network of cosmic strings with $G\mu=10^{-7}$ in a flat $3+1$-dimensional FRW universe, as a function of the scale factor $a$, for various values of $\tilde c$ (see label). The cosmological parameters were set to $\Omega_{\Lambda 0}=0.728$, $\Omega_{r 0} h^2=2.47\times 10^{-5}$ and $h=0.704$, which correspond to the WMAP 7-year data combined with the baryon acoustic oscillation data from the Sloan Digital Sky Survey and determinations of $H_0$ using the Hubble Space Telescope \cite{Jarosik:2010iu}. The scale factor at the time of radiation and matter equality is $a_{eq}$, while its value at the present time was set to unity.}
\label{tran-xi}

\end{center}
\end{figure}

This scaling solution is only attainable for a constant expansion exponent $\beta$ and, therefore, such a regime can only be maintained deep into the radiation or matter eras. Moreover, the values of the scaling constants are different in the radiation and matter epochs. One would then expect the network to experience two different scale-invariant regimes separated by a transitional period (triggered by the radiation-matter transition) during which it is not in linear scaling. This issue has been discussed in detail in \cite{Avelino:2012qy}, and the authors found that this transition is indeed long-lasting, especially for small values of $\tilde c$. This is illustrated in Fig. \ref{tran-xi}, where the cosmological evolution of $\xi$ and $\vv$ is plotted as a function of the scale factor for a cosmic string network in a $3+1$-dimensional FRW universe, for different values of $\tilde c$. Notice that, due to the recent acceleration of the expansion of the Universe, the matter epoch might not be long enough for the network to attain a linear scaling regime before being diluted away by the accelerated expansion.

\section{Stochastic Gravitational Wave Background\label{SGWB}}

Cosmic string loops are copiously produced throughout the cosmological evolution of cosmic string networks. Once these loops detach from the long string network, they oscillate relativistically under the effect of their tension, and decay by emitting gravitational waves. As a result, cosmic string networks are expected to give rise to a stochastic gravitational wave background spanning a wide range of frequencies and having a characteristic shape \cite{Vilenkin:1981bx,Hogan:1984is}.

\subsection{Cosmic String Loop Emission}

In this paper, we shall assume that loops are born with a size which is a fixed fraction of the characteristic length of the string network at that time ($t_b$)
\be
l_b=\alpha L(t_b)\,,
\label{size0}
\ee
where $l=E/\mu$ is the physical length of the cosmic string loops, $E$ is the energy of the loops, $\alpha$ is a constant parameter and the subscript $b$ refers to the instant of cosmic time when the loops were produced. Realistically, one does not expect all the loops produced at a given time to have precisely the same length. Instead, the distribution of the sizes of the loops formed at the time $t_b$ is expected to have a peak around $l_b$. If the width of the distribution of loop sizes is not very large, then assuming that all the loops have the same size at the moment of formation should be a good approximation.
%Note also that loops have, at the moment of formation, peculiar velocities that are subsequently redshifted. Consequently, one should include in Eq. (\ref{size0}), a Lorentz factor of the form $g=\left(1-v_{\ell i}^2\right)^{1/2}$ (where $v_{\ell i}$ is the initial loop velocity) in order to account for this energy reduction. On the remainder of this paper, we shall absorb this factor $g$ into the definition of the parameter $\alpha$. For small loops, it is a good approximation to consider that the expansion of the background has, on average, a negligible impact on their energy. As a consequence, cosmic string loops oscillate with a rms velocity of $\ave{v_\ell^2}\approx 1/2$, and, therefore, $g=2^{-1/2}$\cite{vilenkin2000cosmic}.

Loops emit gravitational waves in a discrete set of frequencies, which are mainly determined by the length of the loop

\be
f_j=\frac{2j}{l}\,,
\label{frdef}
\ee
where $j$ is the harmonic mode and $f_j$ is the corresponding frequency. The energy emitted in gravitational waves by a cosmic string loop in each of the modes is

\be
\frac{dE_{j}}{dt}=P_j G \mu^2\,,
\label{powermode}
\ee
where

\be
P_j=\Gamma j^{-q}/\sum_{m=1}^\infty m^{-q}\,,
\label{powermode2}
\ee
$\Gamma \simeq 65$ \cite{Vilenkin:1981bx,Quashnock:1990wv} is a constant that characterizes the efficiency of the emission mechanism, and $q$ is the spectral index. The value of $q$ depends on the shape of the loops: it should be $q \approx 2$ for kinky loops, while we should have $q\approx 4/3$ for cuspy loops \cite{vilenkin2000cosmic}. There is another effect that needs to be taken into consideration when considering loop emissions: gravitational back-reaction. It has been demonstrated, using both field theory simulations and analytical studies \cite{Battye:1994qa,Battye:1997ji} that the effect of gravitational back-reaction is to damp the higher frequency modes while leaving the lower modes virtually unaffected. A cut-off may, then, be introduced to the summation in Eq. (\ref{powermode2}), so that emission modes with $j>n_s$ are not considered. In this case, Eq. (\ref{powermode2}) remains valid with $\infty$ replaced by an integer cut-off $n_s$. In Sec. \ref{depns}, we shall study the dependence of the SGWB spectrum on $n_s$ and determine the values that this parameter should take.

Cosmic string loops, then, loose energy at a rate

\be
\frac{dE}{dt}=\Gamma G\mu^2\,,
\ee
and, as a consequence, their length decreases. As a matter of fact, we have that

\be
l(t)=\alpha L(t_b)-\Gamma G\mu(t-t_b)\,,
\label{sizeeq}
\ee
for $t>t_b$, assuming, for simplicity, that the loops do not break up into smaller loops due to self-intersection. Loops created in this manner would have a faster decay rate than that of their progenitors and they would emit gravitational waves with larger frequencies.

\subsection{Spectral Density of Gravitational Radiation}

The amplitude of the stochastic gravitational wave background is often quantified by the energy density in gravitational waves, $\rho_{\rm GW}$, per logarithmic frequency interval in units of critical density ($\rho_{\rm crit}$),

\be
\Omega_{\rm GW}=\frac{1}{\rho_{{\rm crit}}}\frac{d\rho_{\rm GW}}{d \log f}\,.
\ee

In order to find the spectral energy density of gravitational waves at a given time $t$, one needs to take into consideration all the contributions from gravitational waves that have a frequency $f$ at the instant $t$, emitted by cosmic string loops created between the time of formation of the network, $t_f$, and $t$. Taking into account the redshift of the gravitational wave frequency ($f\propto a^{-1}$), one has that \cite{vilenkin2000cosmic}

\bq
\frac{d\rho_{\rm GW}}{df}(t)&=&2\pi \int_{t_f}^t dt' \left(\frac{a(t')}{a(t)}\right)^3 \times \\
&\times& \int_0^l dl n(l,t')l h\left(2\pi f l \frac{a(t)}{a(t')}\right)\,,
\eq
where $ n(l,t')dl$ is the number density of cosmic string loops with lengths between $l$ and $l+dl$ at time $t$, and $h$ is a function that describes the spectrum of radiation emitted by a loop.  If one assumes a discrete spectrum (as in Eqs. (\ref{powermode}) and (\ref{powermode2})), this function may be written as \cite{Sanidas:2012ee}

\be
h(z)=G\mu^2\sum_{j=1}^{n_s}P_j\delta\left(z-4\pi j\right)\,,
\ee
where $z=(a(t)/(a(t'))(2\pi f l)$. Therefore, one finds that, at the present time,

\be
\Omega_{\rm GW}(f)=\frac{2 G\mu^2}{\rho_{{\rm crit}0} a_0^5 f}\int_{t_f}^{t_0}dt a^5 (t') \sum_{j=1}^{n_s} j P_j n \left(l_j(t'),t'\right) \,,
\label{omega}
\ee
where the subscript $0$ is used to refer to the value of the corresponding parameter at the present time. Here we have also defined

\be
l_j(t')\equiv\frac{2j}{f}\frac{a(t')}{a_0}\,,
\label{loopsize}
\ee
which is the physical length that the cosmic string loops should have at each instant $t'$ in order to emit, in the harmonic mode $j$, gravitational waves that have a frequency $f$ at the present time. Given this dependence of $l$ on $t'$, the relevant components of the loop distribution function, that may contribute to $\Omega_{\rm GW}$ at a given frequency $f$, are $n(l_j(t'),t')$.

\subsection{Number density of Loops}

In order to compute $\Omega_{\rm GW}$, it is necessary to correctly compute the loop distribution function $n(l_j(t'),t')$. Let $n_c(t)$ be the total number density of loops that have been formed as the result of intercommutation between the time of formation of the string network and time $t$. If one starts by evolving the VOS equations numerically --- in order to compute the characteristic length, $L$, and the RMS velocity, $\vv$, for a discrete set of cosmic times between $t_f$ and $t_0$ --- the rate of loop production (per comoving volume) may easily be determined:

\be
\frac{dn_c}{dt}=\frac{\tilde c}{\alpha}\frac{\vv}{L^4}\,.
\label{nc}
\ee
This expression may simply be obtained by dividing the total energy density that is lost by the string network due to loop formation (Eq. (\ref{loss})) by the energy of each loop at the moment of creation (recall that, for simplicity, all loops produced at a given time are assumed to be created with the same physical length and thus the same energy).

Note that, after formation, loop size shrinks as a consequence of gravitational radiation emission. Therefore, $n(l_j(t'),t')$ has contributions from all preexisting loops that have physical lengths $l_j(t')$ at time $t'$. Determining the times of creation ($t_b^i$) of the loops  that contribute to a given frequency at any given time $t'$ is essential to correctly computing $n(l_j(t'),t')$. Given these instants, one has that

\be
n(l_j(t'),t')=\sum_i n(l_j(t_b^i),t_b^i)\left(\frac{a(t_b^i)}{a(t')}\right)^3\,,
\ee
where we have taken into account the fact that the loops are diluted away by the background expansion. The number density of loops created at the instants $t_b^i$ may be computed using Eq. (\ref{nc}):

\be
n(l_j(t_b^i),t_b^i)=\left. \frac{dn_c}{dt} \right|_{t=t_b^i}\left. \frac{dt}{dl}\right|_{t=t_b^i}\,.
\ee

We then have that

\be
n\left(l_j(t'),t'\right)  =  \sum_i \left\{ \frac{1}{\alpha \left. \frac{dL}{dt}\right|_{t=t_b^i}+\Gamma G\mu} \frac{\tilde c}{\alpha} \frac{\vv(t_b^i)}{L^4(t_b^i)}\left(\frac{a(t_b^i)}{a(t')}\right)^3\right\}\,.
\label{loopdist}
\ee
Note that, in performing this calculation --- contrary to what is generally done in the literature \cite{DePies:2007bm,Binetruy:2012ze,Sanidas:2012ee,Kuroyanagi:2012wm} --- we did not assume the network to be in a linear scaling regime. As we have discussed in Sec. \ref{lin}, once the radiation-matter transition is triggered, for a realistic expansion history and realistic values of $\tilde c$, the cosmic string network is not expected to remain in a scale-invariant regime. As we shall see in the rest of this paper, considering a realistic radiation-matter transition has a significant impact on the shape of the spectrum at low frequencies.

\section{The effect of the radiation-matter transition on the SGWB spectrum\label{transition}}

The stochastic gravitational wave background produced by a cosmic string network is, at any given frequency, the result of the superimposition of the contributions of all the loops that generate,  throughout the cosmic history, gravitational waves with that frequency at the present time. The loops created during the radiation epoch are the dominant contributors to the high frequency range and they produce a flat spectrum. On the other hand, larger loops, created during the matter era, generate a peak in the low frequency portion of the spectrum.

Although the general shape of the SGWB spectrum generated by a cosmic string network is maintained for a wide range of parameters, its amplitude and the characteristics of the peak may vary greatly. In this section, we shall compute the SGWB spectrum generated by cosmic string networks for a wide range of parameters. We perform these computations using two different models for the evolution of cosmic string networks: the VOS model described in Sec. \ref{vos} (Model I) and a model whose underlying assumption is that cosmic string networks experience scale-invariant evolution throughout cosmological history (Model II).  In Model II, the required linear scaling behavior is achieved by determining the values of the scaling parameters during the radiation and matter eras (using Eq. (\ref{sca-c})), and by assuming that $\vv$ and $\xi$ change abruptly between these values, in a step-like manner, at the time of radiation and matter  equality. Note that the approximation made in Model II is quite common in the calculation of the SGWB spectrum generated by cosmic string networks.
Computing the SGWB power spectra for both these models will allow us, not only to assess the differences in the shape of the spectra that result from the variation of cosmic string parameters, but also to study the impact of an inappropriate modeling of the radiation-matter transition on the shape of the spectra.

Throughout this paper, we assume a $3+1$-dimensional FRW background containing matter, radiation and a cosmological constant. The cosmological parameters take the values $\Omega_{\Lambda 0}=0.728$, $\Omega_{r 0} h^2=2.47\times 10^{-5}$ and $h=0.704$ \cite{Jarosik:2010iu}, while $\Gamma=65$. The fiducial value of the energy loss parameter is ${\tilde c}=0.23$.

%study the SGWB spectra generated by VOS cosmic string networks for a wide range of parameters, in order to assess these differences. Furthermore, we will compare these to the spectra generated by cosmic string networks that undergo scale-invariant evolution, in order to understand the impact of the radiation-to-matter transition on the shape of the SGWB spectra. Throughout this section, 

\subsection{Dependency on $\alpha$ and $G\mu$}

The shape and intensity of the SGWB spectrum is affected by the value of the loop size parameter $\alpha$. There is presently no consensus in the literature as to what value should $\alpha$ take. Several studies suggest that $\alpha$ should be smaller than the gravitational back-reaction scale $\alpha < \Gamma G\mu$, while others suggest scales much closer (albeit 1 to 3 orders of magnitude smaller) to the characteristic lengthscale of the network (see \cite{Siemens:2001dx,Siemens:2002dj,Polchinski:2006ee,Siemens:2006yp,Polchinski:2007rg,Copeland:2009dk,BlancoPillado:2011dq,Martins:2005es,Ringeval:2005kr,Vanchurin:2005pa,Olum:2006ix}). There are also studies which indicate that string loops might be formed with lengths similar to the string thickness \cite{Vincent:1996qr,Vincent:1997cx,Bevis:2007gh}, and others suggesting that the loops generated at any given time may be associated  with two or more fundamentally different scales \cite{Vanchurin:2007ee,Lorenz:2010sm}. Given the uncertainties associated to the value of the loop size parameter, we shall compute the SGWB power spectrum for a wide range of values of $\alpha$.

\begin{figure}
\includegraphics[width=3.4in]{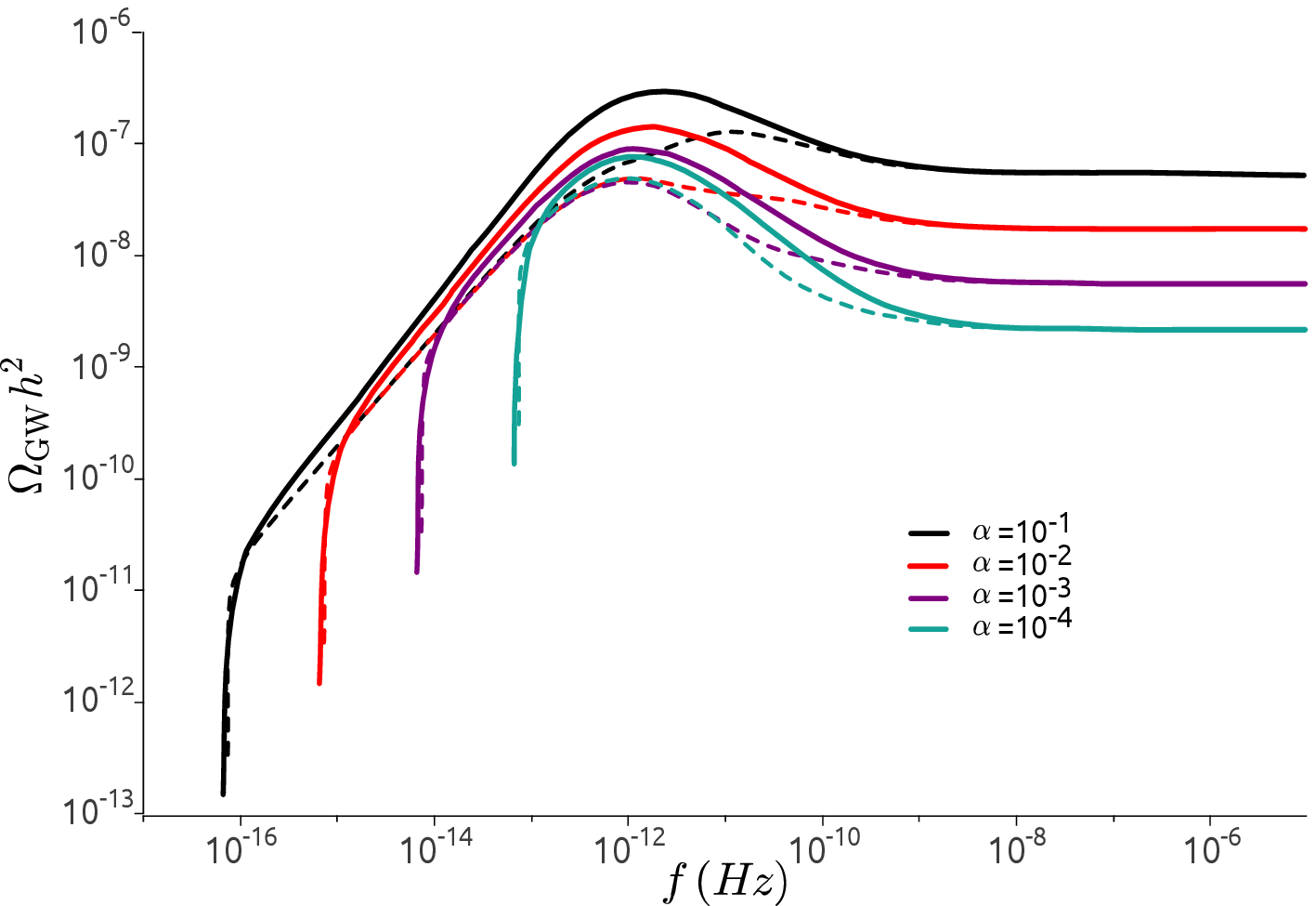}
\caption{The SGWB spectrum, $\Omega_{\rm GW}h^2$, as a function of the frequency, $f$, for various values of $\alpha > \Gamma G\mu$. The solid lines represent the spectra created by a cosmic string network described by the VOS model, while the dashed lines represent the spectra obtained using Model II. Here, we assumed that $G\mu=10^{-7}$ and ${\tilde c}=0.23$, and only the fundamental mode of emission was considered ($n_s=1$).}
\label{lalpha}
\includegraphics[width=3.4in]{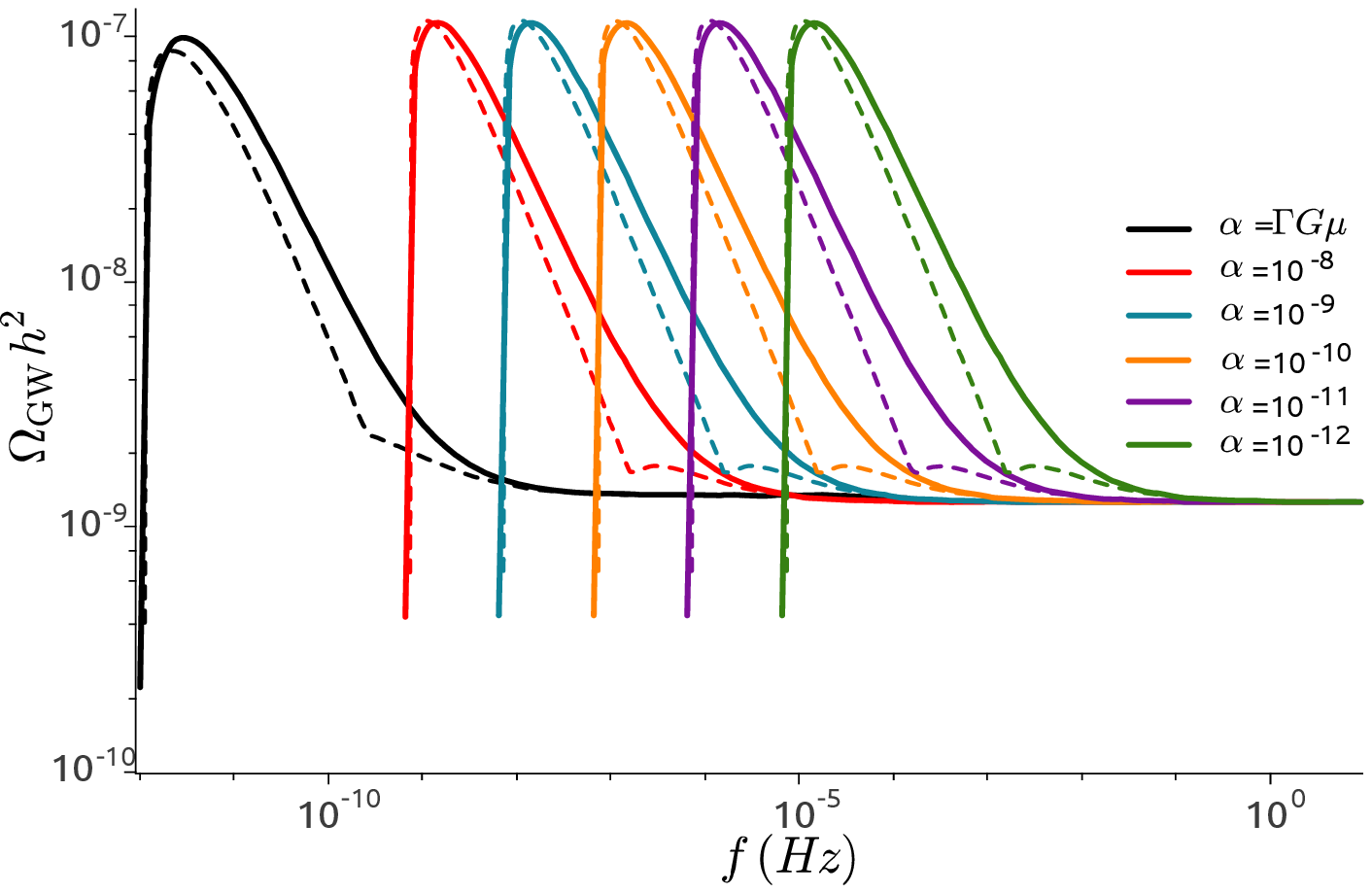}
\caption{The SGWB spectrum, $\Omega_{\rm GW}h^2$, as a function of the frequency, $f$, for various values of $\alpha \le \Gamma G\mu$. The solid lines represent the spectra generated using Model I, while dashed lines represent the spectra obtained using Model II. As in Fig. \ref{lalpha}, we have assumed that $G\mu=10^{-7}$ and ${\tilde c}=0.23$, and $n_s=1$.}
\label{salpha}
\end{figure}

There are two different regimes to take into consideration when analyzing the effects of $\alpha$ on the shape and intensity of the spectrum. Large loops, with sizes above the gravitational back-reaction scale ($\alpha>\Gamma G\mu$), have a lifetime larger than the Hubble timescale ($H^{-1}$). Consequently, in this regime, cosmic string loops persist for a significant amount of time before disappearing completely, emitting gravitational radiation throughout their lifetime. One may then expect the density of gravitational waves to be larger in the case of loops with a larger loop size parameter $\alpha$. This effect is shown in Fig \ref{lalpha}, where the SGWB power spectrum generated using Model I is represented for different values of $\alpha$. Fig. \ref{lalpha} also shows the analogous SGWB power spectra (characterized by the same $G\mu$ and $\alpha$) in the case of Model II. The analysis of the plots shows that Model II predicts lower and narrower peaks, when compared to Model I. This happens because, for realistic values of $\tilde c$, once the radiation-matter transition is triggered, the parameter $\xi$ suffers a slow increase, while $\bar v$ decreases slowly (as is evident in Fig. \ref{tran-xi}). Assuming that this variation occurs abruptly introduces a discontinuity in $n\left(l(t'),t'\right)$ and leads to an underestimation of the number of loops chopped from the long-string network during the matter-dominated era. Additionally, assuming an instantaneous change in the scaling constants causes an overestimation of the size of the loops at the moment of creation during the matter epoch and the consequent overestimation of the lifetime of the loops. Although this has the effect of increasing the peak of the spectra, it is not sufficient, in general, to counteract the other effect described above. At low frequency, there is also a slight difference in the spectra that is caused by the recent acceleration of expansion caused by dark energy. The effect of dark energy is as one might expect: it causes a decrease of the amplitude of the spectrum at low frequencies, because of the decline in the number of loops produced that is caused by the accelerated expansion (during such a phase $L\propto a$ and $\vv\to 0$ \cite{Sousa:2011ew}).

In the small loop regime ($\alpha < \Gamma G\mu$), cosmic string loops have a lifespan smaller than the cosmological time scale and may, therefore, be assumed to radiate their energy immediately after being chopped. It is straightforward to realize that, in this limit, the shape of the spectrum is independent of $\alpha$. However, in this case, the effect of changing $\alpha$ is to change the frequency of the emitted gravitational waves. The spectrum, then, merely suffers a shift towards higher frequencies as $\alpha$ increases, as is clearly illustrated in Fig. \ref{salpha}. As to the differences between the spectra generated using Models I and II, they are mainly caused by the overestimation of $\xi$ during the matter epoch. As matter of fact, a loop created with a size $l$ radiates gravitational energy with frequencies $f\ge 2/l$. Loops created during the matter era by a network that remains artificially in a linear scaling regime during the transitional period would be larger, and, as a consequence, the SGWB spectrum generated by these networks exhibits a shift towards lower frequencies.

\begin{figure}
\includegraphics[width=3.4in]{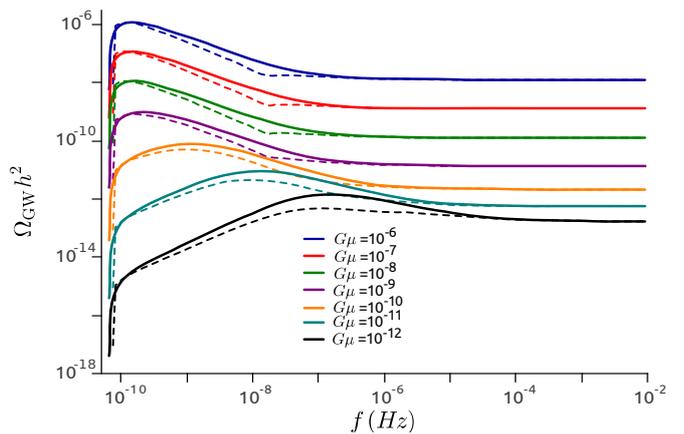}
\caption{The SGWB spectrum, $\Omega_{\rm GW}h^2$, as a function of the frequency, $f$, for various values of $G\mu$. The solid lines represent the spectra generated using Model I, and the dashed lines represent the spectra obtained using Model II. The loop size parameter was set to $\alpha=10^{-7}$, the loop-chopping efficiency was set to ${\tilde c}=0.23$, and only the fundamental mode of emission was considered ($n_s=1$).}
\label{Gmus}
\end{figure}

The effects of varying the string tension, $G\mu$, on the amplitude of the spectra may be explained similarly. First of all, one may once again analyze the large loop and small loop regimes separately. The spectral density depends on $G\mu$ directly (see Eq. (\ref{omega})), but also indirectly since the lifetime of the loops depends on $G\mu$. In the small loop regime, the effect of varying $G\mu$ on the lifetime of the loops is negligible --- since loops radiate their energy effectively immediately after formation. The amplitude of the spectrum, then, varies proportionally to string tension ($\Omega_{\rm GW}h^2\propto G\mu$). If the loops are large, however, the effect of decreasing $G\mu$ is to increase the lifetime of the loop. This causes a less pronounced decrease of the SGWB spectrum amplitude and a shift of the peak towards higher frequencies. These effects are clearly illustrated in Fig. \ref{Gmus}, where the SGWB spectra generated using Models I and II are represented for various values of $G\mu$, and for $\alpha=10^{-7}$. Note also that the differences between Models I and II persist for different values of $G\mu$ and they are quite significant in the mid-to-small frequency range.

\subsection{Dependency on $n_s$ and $q$\label{depns}}

The spectrum of gravitational radiation emitted by cosmic string loops depends both on the cut-off on the number of frequency modes, $n_s$, and the spectral index, $q$. As we shall see in this section, the shape of the stochastic gravitational wave spectrum is highly dependent on the values of these parameters.

\begin{figure}
\includegraphics[width=3.4in]{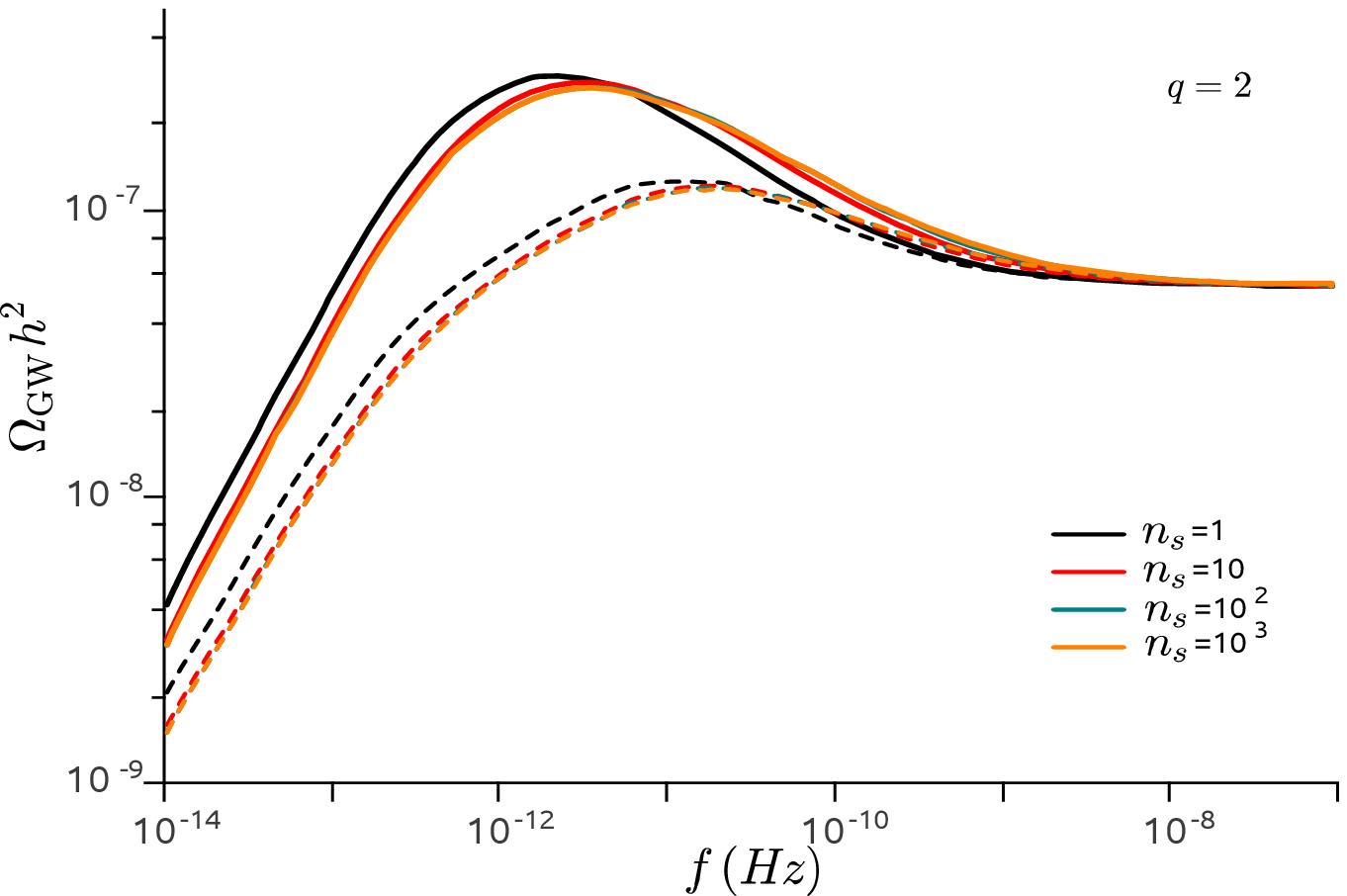}
\caption{The SGWB spectrum, $\Omega_{\rm GW}h^2$, as a function of the frequency, $f$, for different values of $n_s$. The solid lines represent the spectra generated by a cosmic string network described by VOS model (Model I), while the dashed lines are spectra obtained using Model II. Here, we have considered kinky loops (with $q=2$) with a loop size parameter $\alpha=10^{-1}$, and we have assumed that $G\mu=10^{-7}$ and ${\tilde c}=0.23$.}
\label{k1}
\includegraphics[width=3.4in]{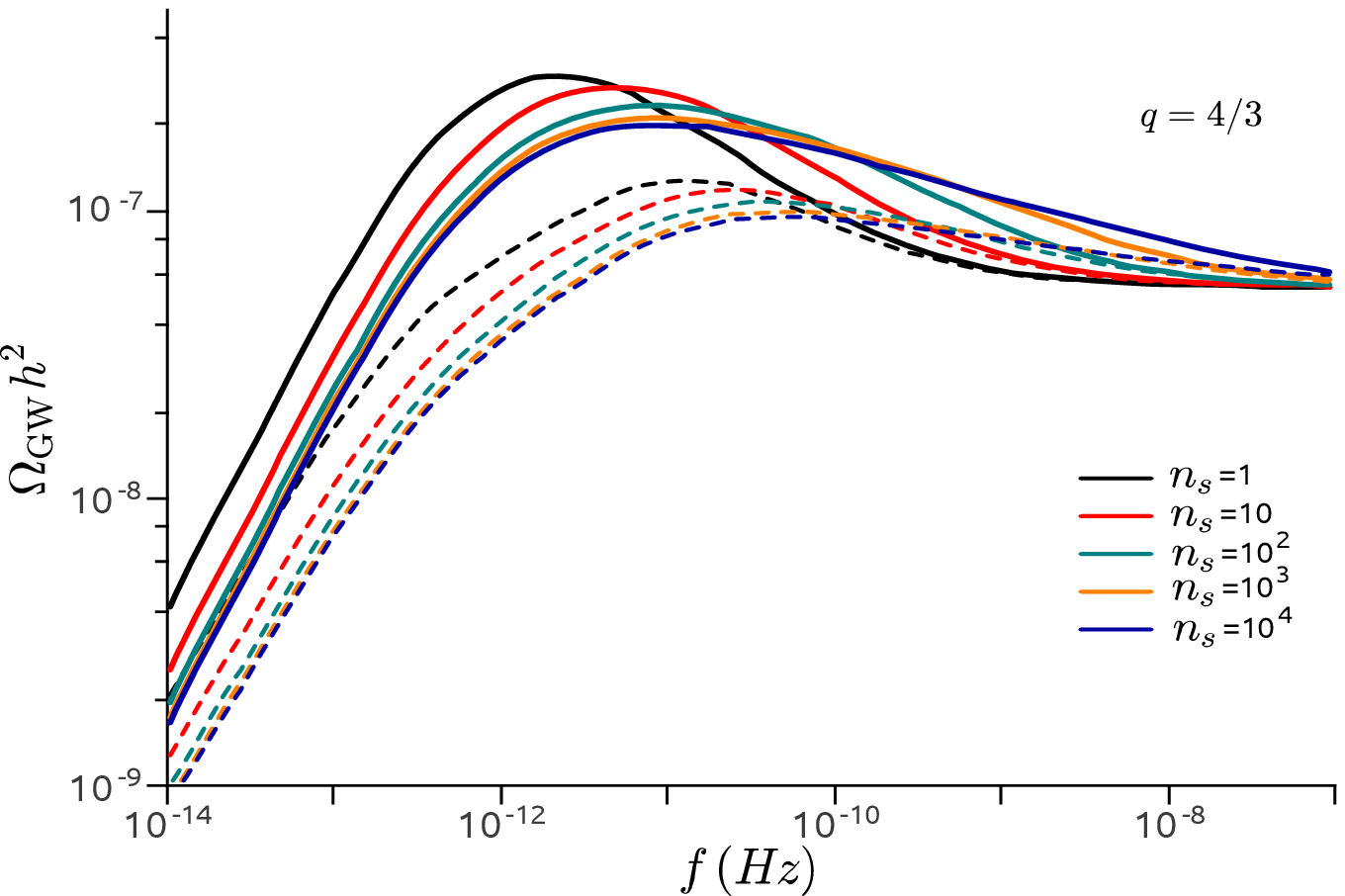}
\caption{The SGWB spectrum, $\Omega_{\rm GW}h^2$, as a function of the frequency, $f$, for different values of $n_s$. The solid lines represent the spectra generated using Model I, while the dashed lines are the spectra obtained using Model II. Here, we considered cuspy loops (with $q=4/3$) with $\alpha=10^{-1}$, and we have assumed that $G\mu=10^{-7}$ and ${\tilde c}=0.23$.}.
\label{c1}
\end{figure}

Figs. \ref{k1} and \ref{c1} show the spectral density of radiation generated by a cosmic string network with a large loop size parameter ($\alpha=10^{-1}$) for different values of $n_s$ and $q$. The inclusion of higher order modes of emission causes a shift of the peak towards higher frequencies. This shift is, in both cases, accompanied by a decrease of the maximum amplitude of the spectrum and by a broadening of the peak. These effects may easily be inferred from Eqs. (\ref{powermode}) and (\ref{powermode2}): they are simply a consequence of weighing in higher order frequency modes. One might also deduce from these expressions that the spectrum generated by cosmic string networks with cuspy loops ($q=4/3$) is significantly more affected than that of string networks with kinky loops ($q=2$). In both cases, these differences are more significant for larger values of $n_s$. However, one may notice, by analyzing Figs. \ref{k1} and \ref{c1}, that the spectra seem to converge to a fixed shape for large values of $n_s$. Although in the case of kinky loops it is sufficient to consider modes up to $n_s=10^2$, for cuspy loops the cut-off should be somewhat larger ($n_s=10^4$) (these results seem to be in agreement with those found in Ref. \cite{Sanidas:2012ee}).

\begin{figure}
\includegraphics[width=3.4in]{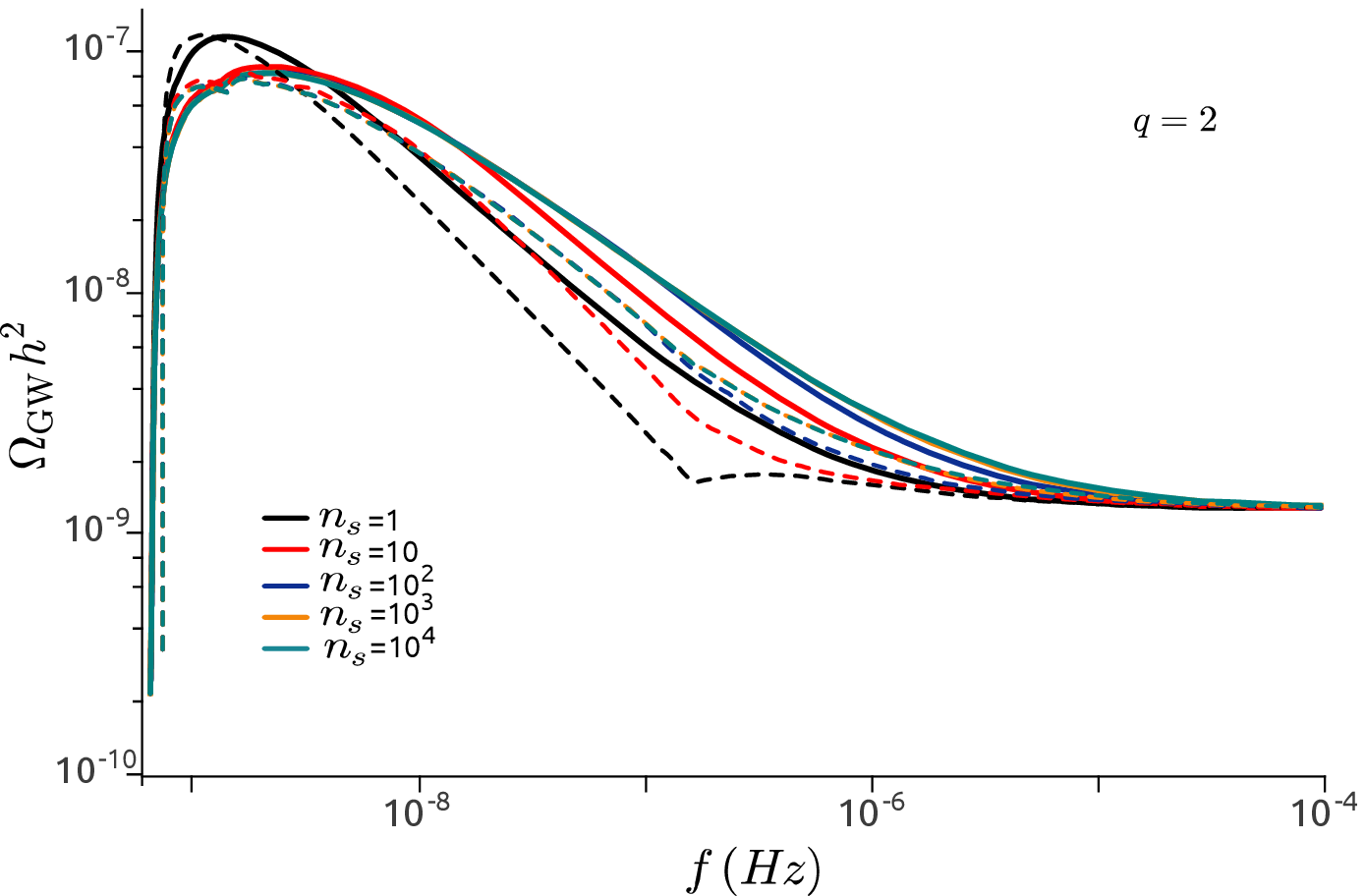}
\caption{The SGWB spectrum, $\Omega_{\rm GW}h^2$, as a function of the frequency, $f$, for different values of $n_s$. The solid lines represent the spectra generated by a cosmic string network described by the VOS model (Model I), while the dashed lines are obtained using Model II. Here, we considered kinky loops (with $q=2$) with a loop size parameter $\alpha=10^{-8}$, and we have assumed that $G\mu=10^{-7}$ and ${\tilde c}=0.23$.}
\label{k8}
\includegraphics[width=3.4in]{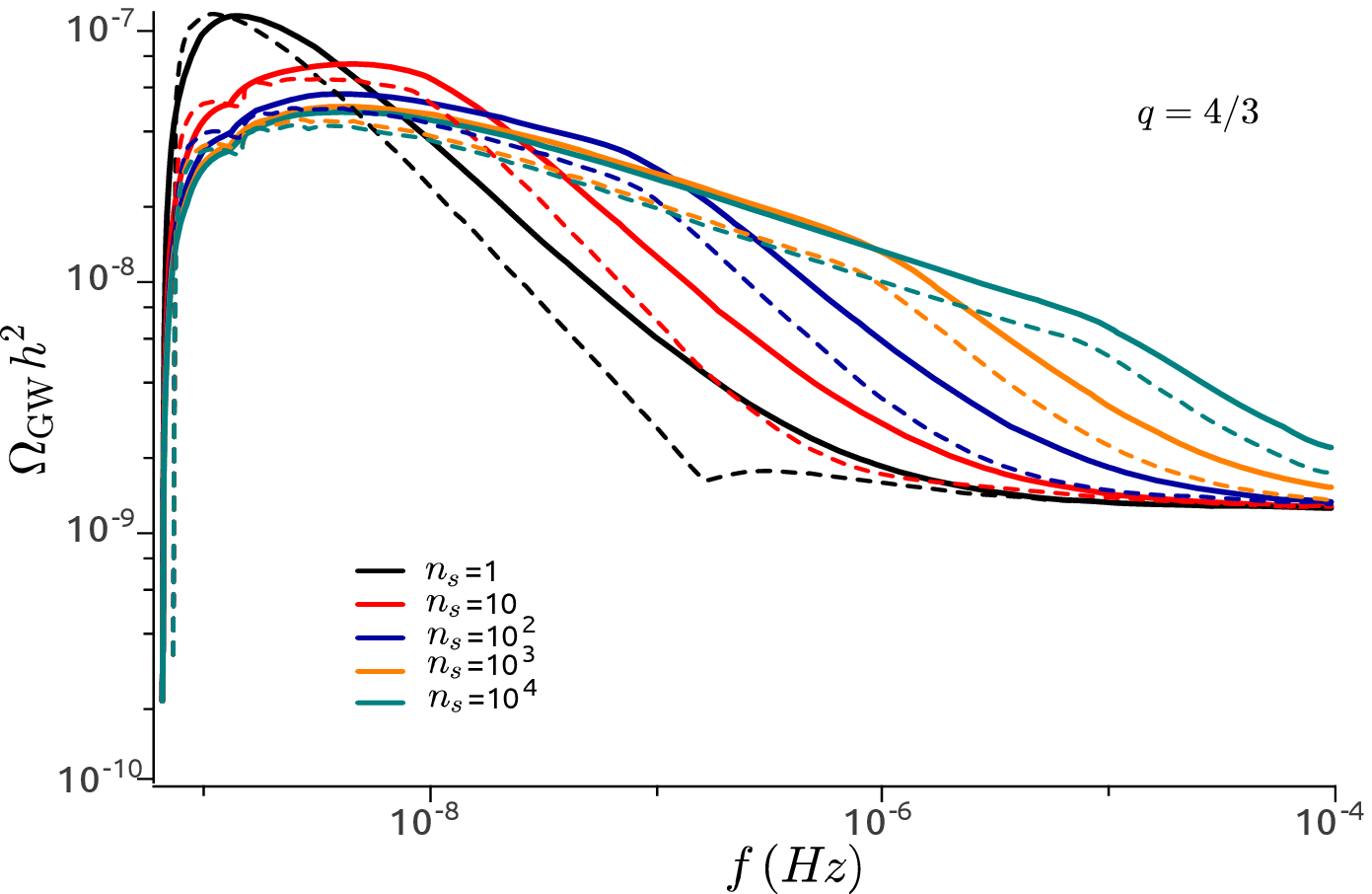}
\caption{The SGWB spectrum, $\Omega_{\rm GW}h^2$, as a function of the frequency, $f$, for different values of $n_s$. The solid lines represent the spectra generated using Model I, while the dashed lines are obtained using Model II. Here, we considered cuspy loops (with $q=4/3$) whose size is determined by $\alpha=10^{-8}$, and we have assumed that $G\mu=10^{-7}$ and ${\tilde c}=0.23$.}.
\label{c8}
\end{figure}

When one considers a small loop size parameter, the picture is somewhat changed. One may see in Figs. \ref{k8} and \ref{c8} that the effect of including higher order modes is qualitatively the same. However, the shape of the spectrum is significantly affected for cuspy loops ($q=4/3$). In this case, the broadening of the peak is very significant and the spectrum develops a hump as it decreases towards the constant portion. Moreover, for $\alpha<\Gamma G\mu$, one needs to consider values of $n_s$ that are about one order of magnitude higher to achieve convergence of the shape of the spectrum. Notice also that the spectra exhibits small "bumps" in the low frequency range. These are a consequence of the fact that loops emit gravitational waves in a discrete set of frequencies. Recall that the gravitational waves emitted by the $j$-th harmonic have a minimum frequency of $f=jf_0$ (where $f_0$ is the frequency of gravitational waves emitted at the present time in the fundamental mode) and, thus, this mode does not contribute to the spectral amplitude of gravitational radiation for $f<jf_0$. Each successive harmonic comes into play at $f=jf_0$, thus causing a sudden change in the slope of the SGWB spectrum. Note that these discontinuities are more evident for a small loop size parameter because, in this case, the values $f=jf_0$ coincide with the peak region (whose slope is quickly varying). Note also that these "bumps" would not be present if one would consider the more realistic case of the loop size following a distribution of values around $\alpha L(t_b)$ (instead of considering only one loop size).

By analyzing Figs. \ref{k1} to \ref{c8}, one may conclude that the differences between the spectra generated using Models I and II are still present for large values of $n_s$ and for $q=2$ (kinky loops) and $q=4/3$ (loops with cusps). For large values of $\alpha$, the differences are essentially of the same magnitude as in the case of $n_s=1$. For small $\alpha$, however, the maximum amplitude of the spectrum generated using Model I surpasses that obtained using Model II for the $n_s>1$ case (despite the fact that the amplitude of the peaks was very similar for $n_s=1$). This happens because, in the case of Model II, the peak is situated at very low frequencies, close to $f_0$. This is the region that suffers a more accentuated decrease as a result of considering larger values of $n_s$.

\subsection{Dependency on $\tilde c$}

String-theory-motivated scenarios suggest that fundamental strings (F-strings) and $1$-dimensional Dirichlet branes (D-strings) may grow to macroscopic scales and play the cosmological role of ordinary cosmic strings (see e.g. \cite{Dvali:2001fw,Jones:2003da,Copeland:2003bj}). These cosmic superstrings --- unlike ordinary cosmic strings --- have a reconnection probability, $P$, that may be significantly smaller than unity: for instance, for D-string intersections $0.1 \lesssim P \lesssim 1$, while for collisions between F-strings the expectation is $10^{-3}\lesssim P\lesssim 1$ \cite{Jackson:2004zg}. Hence, in D-D or F-F collisions, there is a non-vanishing probability of the strings merely passing through each other without interaction.

It is straightforward to realize that, as a result of the reduction of the reconnection probability, the amount of loops created throughout the evolution of the network is also significantly reduced. The energy-loss mechanism of these networks is, thus, less efficient, and  the loop chopping parameter is expected to scale as $\tilde c \propto P$. For a fixed value of $G\mu$, weakly interacting cosmic superstring networks have, as a consequence, a larger string energy density and thus an enhanced SGWB amplitude.

As a matter of fact, during the radiation era, weakly interacting string networks with ${\tilde c}\ll 1$ attain a linear scaling regime characterized by \cite{Avelino:2012qy}, 
\be
\xi=\sqrt{2} {\tilde c}\,,\qquad \bar{v}= \frac{1}{\sqrt{2}}-\delta\,,
\ee
with $\delta=(\pi/12){\tilde c}$. Using Eqs. (\ref{size0}-\ref{frdef}) and (\ref{loopdist}), we may then conclude that, for weakly interacting networks, $n(l,t) \propto P^{-3}$ and $f \propto P^{-1}$, during the radiation era. As a result, the amplitude of the flat portion of the SGWB spectrum should scales as

\be
\Omega_{\rm GW}\propto P^{-2}\,.
\label{lowp}
\ee

As to the matter era, the fact that the network is not experiencing scale invariant evolution makes quantitative analysis more difficult. Fig. \ref{tran-xi} shows that the effect of varying $\tilde c$ on $\vv$ and $\xi$ is non trivial. On the one hand, for ${\tilde c}\ll 1$, the scaling law $\xi \propto P$ is approximately maintained. However, for larger values of $\tilde c$, the increase in $\xi$ is somewhat smaller and, as consequence, the decrease on the number of loops created that occurs as a result of the radiation-matter transition is less accentuated. Consequently --- although for very weakly interacting networks ($P\ll 1$) Eq. (\ref{lowp}) is still valid --- for $P \sim 1$ the decrease of the amplitude of the peak of the SGWB spectrum is less accentuated.

\section{Conclusions\label{con}}
A cosmic string network is not expected to maintain a linear scaling regime throughout the cosmic history. During the radiation-matter transition, the network needs to adapt to the changing background conditions, entering a phase in which the parameters $\xi$ and $\vv$ are not constant. Moreover, the matter epoch is not long enough for the network to be able to attain scale-invariant evolution before the universe becomes effectively dark-energy-dominated. Therefore, it is not realistic to assume that cosmic string networks remain in a scaling regime after the transition between the radiation- and matter-dominated eras.

Most of the computations of the SGWB power spectrum available in the literature assume that the network experiences linear scaling evolution, with the scaling parameters changing abruptly at the time of radiation-matter equality. In this work, we have assessed this assumption by studying in detail the spectra generated by cosmic string networks that undergo a realistic cosmological evolution. This was done for a wide range of parameters and the results were compared with those of models in which the strings are assumed to remain in a linear scaling regime. We found that the linear scaling assumption has significant impact on the shape and amplitude of the spectrum, in particular, in the low-to-mid frequency range. This range corresponds to the peak of the spectrum which has, in general, an amplitude that is a few orders of magnitude larger than that of its flat portion, and may then be within the reach of future experiments. 

\begin{figure}
\includegraphics[width=3.4in]{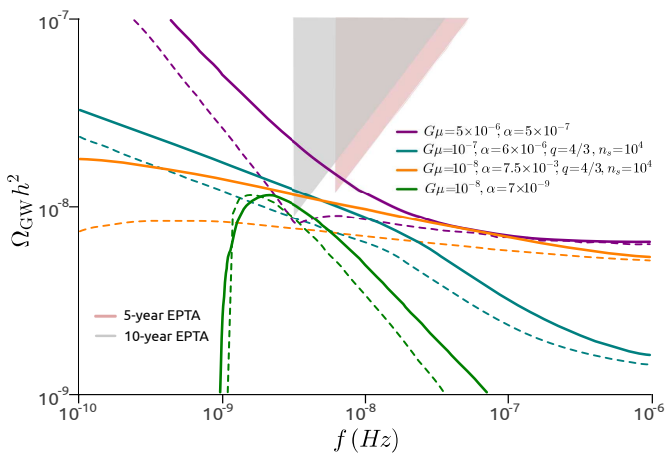}
\caption{Some examples of SGWB spectra generated by cosmic string networks (see label for details) that would not seem to be within the detection range of the 5-years (pink area) and 10-years (gray area) European Pulsar Timing Array (EPTA) experiments \cite{vanHaasteren:2011ni}, when scale invariance is assumed (dashed lines). In these examples, the equivalent spectra obtained using the VOS model (solid lines) are within the detention range of the EPTA experiments.}
\label{obs}
\end{figure}

In Fig. \ref{obs}, some examples of SGWB spectra generated using Model I that are within the sensitivity range of the European Pulsar Timing Array (EPTA) experiments \cite{vanHaasteren:2011ni} are plotted. In all these examples, the spectra obtained using Model II would not seem to be detectable. This clearly illustrates the importance of accurately modeling cosmic string network dynamics, when computing the Stochastic gravitational wave background. The power spectra computed under the assumption that the cosmic string networks remain in a linear scaling regime have narrower peaks, characterized by a lower amplitude and a larger slope, for a wide range of parameters. This implies that simplified models of string evolution may lead to inaccuracies in the computation of the constraints on cosmic string tension $G\mu$ using gravitational wave detection experiments. In particular, for those experiments that have a sensitivity window in the low frequency range, the constraints obtained using the linear scaling assumption are weaker than those obtained using the VOS model for the evolution of the cosmic string network. 

\begin{acknowledgments}
L.S. is supported by Funda\c{c}\~{a}o para a Ci\^{e}ncia e Tecnologia (FCT, Portugal) and by the European Social Fund (POPH/FSE) through the grant SFRH/BPD/76324/2011. P.A. and L.S. are also partially supported by grant PTDC/FIS/111725/2009 (FCT).
\end{acknowledgments}

\bibliography{SGWB}
\end{document}